\documentclass[sigplan,10pt,nocopyright]{acmart}

\acmConference[Off the Beaten Track 2018]{Off the Beaten Track}{January 13, 2018}{Los Angeles, CA, USA}
\acmYear{\vspace{-1em}}
\acmISBN{} 
\acmDOI{} 
\startPage{1}

\setcopyright{none}

\usepackage{booktabs}   
\usepackage{subcaption} 

\usepackage{xcolor}

\begin{document}

\title{Synthesizing Program-Specific Static Analyses}

\author{Colin S.\ Gordon}
\orcid{0000-0002-9012-4490}
\affiliation{
    \institution{Drexel University}
}
\email{csgordon@drexel.edu}

\begin{abstract}
Designing a static analysis is generally a substantial undertaking, requiring significant expertise in both program analysis and the domain of the program analysis, and significant development resources.  As a result, most program analyses target properties that are universallly of interest (e.g., absence of null pointer dereference) or nearly so (e.g., deadlock freedom).  However, many interesting program properties that would benefit from static checking are specific to individual programs, or sometimes programs utilizing a certain library.  It is impractical to devote program analysis and verification experts to these problems.

We propose instead to work on example-based synthesis of program analyses within well-understood domains like type qualifier systems and effect systems.
The dynamic behaviors behind the classes of problems these systems prevent correspond to examples that developers who lack expertise in static analysis can readily provide (data flow paths, or stack traces).
\end{abstract}

\maketitle

\section{Introduction}

Program analysis and verification, broadly, has many examples of successful tools in both research and practice.  Program analysis is now capturing the attention of practicing software developers, whose interests lay primarily in producing functioning software in a timely fashion.
The area spans the gamut from high-end concurrent program logics capable of verifying the details of the most sophisticated algorithms, to more modest properties like absence of null pointer dereferences~\cite{dietl11}, freedom from standard concurrency bugs (e.g., data races~\cite{rccjava00}), or simply performing UI updates on the thread expected by the user interface library~\cite{ecoop13}.
For all points along this spectrum, however, the designers of these analyses justify their efforts in terms of impact: the substantial effort involved in developing each of these type systems or program logics is worthwhile because the effort applies to nearly all software written in a given language.

We view this reasoning as slightly misguided: while it certainly identifies problems with strong potential for impact, it is too narrow to justify other worthy endeavors.  Many properties practicing developers would be interested in checking statically are specific to a particular system, team, or internal library.  Research to statically check these properties will never be justified by the common reasoning, but static checking for them may still be deeply valuable.  Large companies including Microsoft, Facebook, Uber, and Samsung can afford to fund and staff dedicated teams that build analyses for the specific needs of specific projects (and have done so).  But only a small fraction of companies can manage this, and even within those companies only a fraction of developers reap the benefits.

There is a different way to have broad impact through program analysis, without restricting attention to specific problems with broad appeal.  Instead, we can shift some effort towards \emph{classes} of problems, of which many instances may exist, even if individual instances have relatively small potential user bases.  Then we can focus on finding ways to derive instances of these classes without program analysis or verification expertise.
We have already laid some groundwork for this: there already exist generic characterizations and implementations of type qualifiers~\cite{Foster1999,dietl11}, effect systems~\cite{marino09,tate13,ecoop17}, and abstract interpreters~\cite{cousot1977abstract}, among others --- we can use developer examples to infer the instantiations of these frameworks.

In the remainder of this paper, we give two examples of such classes of systems, and show how types of examples \emph{familiar to practicing developers} can be used to infer an analysis from a known class.  We close by discussing broader challenges.

\section{Synthesizing Type Qualifiers}
Type qualifiers~\cite{Foster1999} are a well-established technique for constraining propagation of data, by attaching an extra marker to types that classifies some intrinsic or extrinsic property of the classified data.  This qualifier then also participates in subtyping according to a partial order on the qualifiers.  For example, \textsf{untainted} data could be treated as \textsf{tainted}, or \textsf{nonnull} references could be treated as \textsf{nullable} --- but not the other direction in either example.

In each such system, the partial order on qualifiers amounts to a restriction on which kinds of data can flow to uses of which other kinds of data --- a restriction on data flow.  
This is a familiar concept to practicing developers, who already debug issues with inappropriate data flows, from SQL injections to null pointer dereferences to bugs from mixing up different string formattings of the same concepts.  And these are problems for which manually-defined type qualifier systems already work well~\cite{dietl11}.

But each of these systems corresponds, primarily, to a choice of qualifiers and a partial order on them.  We should be able to infer a set of qualifiers and a partial order from a set of negative examples of data flows --- examples of a piece of data making its way from one place to another location it should not reach in a program.  Given a number of dynamic traces of such prohibited data flows (generated by hypothetical dynamic instrumentation built for our purpose), preventing the illegitimate flows corresponds to finding the minimum number of edges (between source-level bindings) to cut in the dynamic flows to break those flows.  Additional constraints may be relevant (for example, some edges may be known to be required, possibly based on a larger number of positive examples from trusted runs).  
This is not a new problem for the PL community: finding a minimum number of partial order constraints to ignore corresponds to one way of localizing type inference errors, for example~\cite{Loncaric2016}.
The rejected edges hint at the partial order among qualifiers, so what remains is to find a way to cluster a larger set of possible (e.g., observed) data flows while keeping vertices from rejected flows in different clusters.
This is already similar to some partial order reduction techniques~\cite{basten2001enhancing}.

\section{Effect Systems From Stack Traces}
\emph{Commutative} effect systems (the traditional style that discards program order information) are ultimately restrictions on the context in which certain actions may occur, whether this describes Java's checked exceptions, or any number of other systems.  This means that occurrences of the bugs such effect systems prevent are described by a stack trace, which includes at least one instance of an operation with a larger effect (higher in the join semilattice) that occurs during the dynamic extent of some unit of code (function, etc.) whose intended effect would be smaller (lower in the semilattice).

Stack traces are familiar to nearly all practicing developers, through their use in debuggers or in diagnostics from exceptions.  Some problems amenable to effect systems even naturally yield stack traces when developers violate the (implicit) effect discipline.  For example, Gordon et al.~\cite{ecoop13} give an effect system for ensuring updates to UI elements are run on a distinguished UI event loop thread.  This threading discipline is mandated and dynamically enforced by most GUI frameworks: calling methods on most UI elements from the wrong thread leads to an exceptions --- which contains a negative stack trace.
Android has a similar intended thread confinement discipline\footnote{\url{https://developer.android.com/studio/write/annotations.html\#thread-annotations}}, also enforced via dynamic checks and exceptions.
Modifying a debug build to produce negative examples for other suitable problems given a reproducible bug would be straightforward.

These examples of illegitimate stack traces then form the same type of example as the data flow paths in the qualifier case.  Finding a minimum number of edges to prohibit statically (subject to other constraints) corresponds to finding a join semilattice of effects.

\section{Challenges, and Looking Forward}

The quality of the semilattices inferred for the approaches above depends heavily on having a useful set of bad examples --- more examples constrains the minimum choice further, and underconstrained instances may lead to nonsensical choices of edges to prohibit.
In some cases, it may be difficult to produce enough exemplar stack traces or data flows to yield a good solution.  In this case it might be possible to supplement with developer-chosen endpoints, with paths filled in from a static callgraph or points-to analysis, taking all appropriately directed paths from source to sink as candidate bad paths.  This is sensible --- the precursor to Gordon et al.'s effect system for thread confinement was in fact an analysis on paths through a callgraph~\cite{ZhangLE2012}.  But the overapproximation may be problematic.

Polymorphism is another challenge, because it will not be explicit in data flows or stack traces.  It seems likely that templates can be used to recognize common forms of polymorphism, as has been done for trace-based type inference~\cite{ecoop16}.  More broadly, this corresponds to disallowing \emph{paths} through data/control-flow graphs, rather than edges.

Type qualifiers and effect systems are only two examples of program analyses with generic characterizations that could in principle be inferred from some kind of examples that nearly any developer could provide.  They have nearly the same abstractions (join semilattices), but this could eventually work for more sophisticated classes like sequential effect systems~\cite{tate13,ecoop17} or abstract interpretation~\cite{cousot1977abstract}.

\end{document}